\documentclass[twocolumn,showpacs,preprintnumbers,amsmath,amssymb,superscriptaddress]{revtex4-1}

\usepackage{graphicx}
\usepackage{dcolumn}
\usepackage{bm}
\usepackage[version=3]{mhchem} 
\usepackage{amsmath}

\usepackage{siunitx}
\usepackage{wasysym}
\usepackage{array}
\usepackage{graphicx}
\usepackage{dcolumn}
\usepackage{bm}
\usepackage{color}
\usepackage{ulem}
\usepackage{pifont}
\usepackage{natbib}
\usepackage{hyperref}
\hypersetup{
colorlinks = true,
urlcolor   = blue,
linkcolor  = blue,
citecolor  = blue
}
\usepackage{nicefrac}

\newcommand{\nmo}{\ce{NaMn2O4}}
\newcommand{\cfo}{\ce{CaFe2O4}}
\newcommand{\lmo}{\ce{Li_{0.92}Mn2O4}}
\newcommand{\cc}{\textit{c}}
\newcommand{\x}{\textit{x}}
\newcommand{\y}{\textit{y}}
\newcommand{\ac}{\textit{a}}
\newcommand{\bc}{\textit{b}}

\newcommand{\Vc}{\textit{V}}

\newcommand{\Pnam}{\textit{Pnam}}
\newcommand{\doc}{$^{\circ}$C}
\newcommand{\dd}{$^{\circ}$}
\newcommand{\TN}{$T_{\rm N}$}
\newcommand{\TTN}{$T_{\rm N1}$}
\newcommand{\TTTN}{$T_{\rm N2}$}
\newcommand{\musr}{$\rm \mu^+SR$}

\begin{document}

\preprint{APS/PRR}
\title{Neutron powder diffraction study of \nmo~and \lmo:\\
New insights on spin-charge-orbital ordering}

\author{N.~Matsubara}
 \email{namim@kth.se}
\author{E.~Nocerino}
\affiliation{Department of Applied Physics, KTH Royal Institute of Technology, SE-10691 Stockholm, Sweden}
\author{K.~Kamazawa}
\affiliation{Neutron Science and Technology Center, Comprehensive Research Organization for Science and Society (CROSS), Tokai, Ibaraki 319-1106, Japan}
\author{O.~K.~Forslund}
\affiliation{Department of Applied Physics, KTH Royal Institute of Technology, SE-10691 Stockholm, Sweden}
\author{Y.~Sassa}
\affiliation{Department of Physics, Chalmers University of Technology, Gothenburg SE-41296, Sweden}
\author{L.~Keller}
\affiliation{Laboratory for Neutron Scattering and Imaging, Paul Scherrer Institut, CH-5232 Villigen PSI, Switzerland}
\author{V.~V.~Sikolenko}
\affiliation{Joint Institute for Nuclear Research, 141980 Dubna Russia}
\affiliation{Karlsruhe Institute of Technology, 76131 Karlsruhe Germany}
\author{V.~Pomjakushin}
\affiliation{Laboratory for Neutron Scattering and Imaging, Paul Scherrer Institut, CH-5232 Villigen PSI, Switzerland}
\author{H.~Sakurai}
\affiliation{National Institute for Materials Science, Namiki 1-1, Tsukuba, Ibaraki 305-0044, Japan}
\author{J.~Sugiyama}
\affiliation{Neutron Science and Technology Center, Comprehensive Research Organization for Science and Society (CROSS), Tokai, Ibaraki 319-1106, Japan}
\author{M.~M{\aa}nsson}
 \email{condmat@kth.se}
\affiliation{Department of Applied Physics, KTH Royal Institute of Technology, SE-10691 Stockholm, Sweden}


\date{\today}

\begin{abstract}
The high-pressure synthesised quasi-one-dimensional compounds \nmo~and \lmo~are both antiferromagnetic insulators, and here their atomic and magnetic structures were investigated using neutron powder diffraction. The present crystal structural analyses of \nmo~reveal that Mn$^{3+}$/Mn$^{4+}$ charge-ordering state exist even at low temperature (down to 1.5~K). It is evident from one of the Mn sites shows a strongly distorted \ce{Mn$^{3+}$} octahedra due to the Jahn-Teller effect. Above \TN~=~39~K, a two-dimensional short-range correlation is observed, as indicated by an asymmetric diffuse scattering. Below \TN, two antiferromagnetic transitions are observed (i) a commensurate long-range Mn$^{3+}$ spin ordering below 39~K, and (ii) an incommensurate Mn$^{4+}$ spin ordering below 10~K. The commensurate magnetic structure ($k_{\rm C} = 0.5, -0.5, 0.5$) follows the magnetic anisotropy of the local easy axes of Mn$^{3+}$, while the incommensurate one shows a spin-density-wave order with $k_{\rm IC} = (0, 0, 0.216)$. For \lmo, on the other hand, absence of a long-range spin ordered state down to 1.5~K is confirmed. 
\end{abstract}

\keywords{\cfo~ structure, Jahn-Teller orbital ordering, magnetic structure}

\maketitle

\newpage
\section{\label{sec:Intro}Introduction}
Low dimensional and frustrated spin systems have attracted attention in experimental as well as theoretical research fields \cite{Masuda2006, Mihaly2006}. In particular, the compounds with a calcium-ferrite \cfo~structure (CFO), so-called ``post spinels", have been extensively studied in the last decade because of the complexity of their competing interactions on a magnetic arrangement \cite{Ling2001, Nozaki2010, Damay2010, Ling2013a, Liu2014a, Awaka2005}. The crystal structure of CFO is based on edge-sharing Fe octahedra chains running along the \cc-axis, and are connected with each other by corner-sharing in the \textit{ab}-plane. These six chains builds up distorted hexagonal void ('tunnels'), where Ca ions are situated [see Fig.~\ref{structfig}]. They pose a geometrically frustrated lattice, based on a honeycomb like mesh of triangular or zigzag ladders of magnetic ions. Such a frustrated system is known to show intriguing magnetic ground states and physical properties, \textit{e.g.} spin-glass, spin-liquid and multiferroicity \cite{Khomskii2009, Kawamura1998}. In fact, for CFO-\ce{NaCr2O4}, which was reported to exhibit an unconventional colossal magnetoresistance effect below N\'eel temperature (\TN~$\sim$ 125~K) \cite{Sakurai2012, Taguchi2017}, a combined work with positive muon spin rotation and relaxation (\musr) and neutron powder diffraction (NPD) has clarified the presence of two-dimensional (2D) antiferromagnetic (AFM) coupling among one-dimensional (1D) zigzag \ce{Cr2O4} chains\cite{Miwa2017, Higuchi2015}.

Furthermore, a tunnel structure along the \cc-axis in the CFO lattice provides an 1D conduction pathway for cations [see the structure in Fig.~\ref{structfig}(a)]. Hence, the CFO compounds with Li$^+$ and Na$^+$ ions are expected to show high ionic conductivity. This makes the CFO compounds attractive for the use as solid state electrolyte material in future all-solid-state batteries \cite{Mukai2019, Ling2013a, Mamiya2013, Mamiya2016}.      

When Li$^+$ and Na$^+$ occupy the $A$ site in a CFO structure represented by $AB_2$\ce{O4} ($A$ = Li and Na, $B$ = Mn, Cr and V), the remaining $B$ ions should be in a mixed valance state with $B^{3+}$ and $B^{4+}$. According to the previous work on CFO-\ce{NaV2O4} \cite{Yamaura2007}, \ce{NaCr2O4} \cite{Sakurai2012} and \lmo~\cite{Yamaura2006, Mukai2019}, there are no indication for charge-ordering in the zigzag chains. Instead, trivalent and tetravalent cations occupy the octahedral $B$ site randomly to keep local charge neutrality. However, \nmo~at room-temperature exhibits the Mn$^{3+}$/Mn$^{4+}$ charge ordering, evidenced from undistorted Mn$^{+4}$\ce{O6} and a strongly distorted Mn$^{+3}$\ce{O6} due to the Jahn-Teller (JT) effect \cite{Akimoto2009, Awaka2005}. 


 \begin{figure}[ht]
   \begin{center}
     \includegraphics[width=\linewidth]{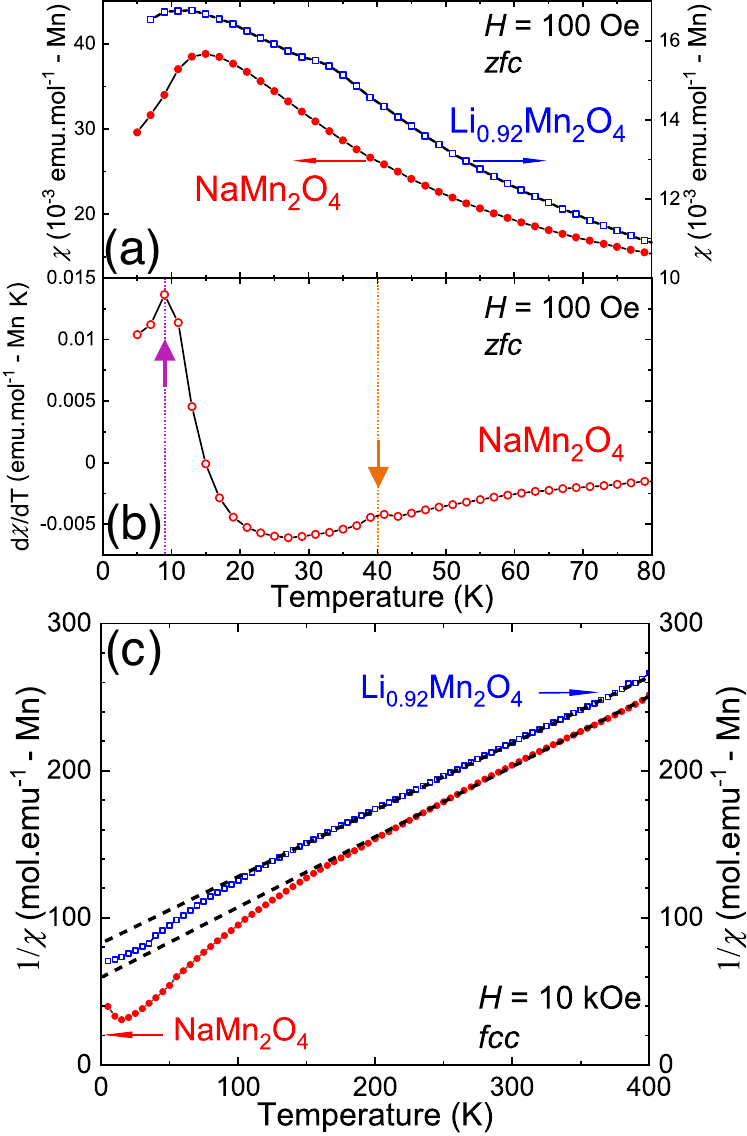}
   \end{center}
   \caption{(a) Magnetic susceptibility curves ($\chi$ ($T$)) of \nmo~(red filled circle: left axis) and \lmo~(blue square: right axis) recorded using a zero-field-cooled warming (zfcw) protocol under an external magnetic field of 100~Oe and (b) of the corresponding differential susceptibility [$d\chi$/$dT$]($T$) curve indicating two magnetic transitions. (c) 1/$\chi$($T$) with Curie-Weiss fitting (dotted lines) recorded a field-cooled cooling (fcc) under an external magnetic field of 10~kOe.}
   \label{sus}
 \end{figure}

The magnetic properties of both \nmo~and \lmo~were previously studied with magnetization and  \musr~measurements \cite{Sugiyama2009b}. The magnetization-vs.-temperature [$M$($T$)] curve for \nmo~[Fig.~\ref{sus} (a)] exhibits a clear maximum at 15~K, indicating the presence of an AFM transition, while the $M$($T$) curve for \lmo~shows a weak anomaly below 40~K. As seen in Fig.~\ref{sus} (c), the inverse susceptibility curves show a large linear domain, for which Curie-Weiss fitting leads to $\mu_{\rm eff}$ and $\Theta_{\rm p}$ values: $\sim$3.35 $\mu_{\rm B}$/Mn and $\sim$-49.6~K for \nmo~and $\sim$4.00 $\mu_{\rm B}$/Mn and $\sim$-161~K for \lmo~\cite{Sugiyama2009b}.   

The \musr~measurements under a weak transverse field revealed that \TN~=~39~K for \nmo~and 44~K for \lmo. However, the zero-field \musr~spectrum lacked a clear oscillation caused by the formation of a static internal magnetic fields for both \nmo~and \lmo. This suggests that the internal AFM field is too rapidly fluctuating and the muons are unable to \textit{see} it, i.e. out of the range of \musr~time window. Therefore, in order to further study the AFM nature in \nmo~and \lmo, the neutron diffraction technique is necessary, particularly for clarifying the difference between \nmo~and \lmo.

In this article, we report a systematic NPD study of both \nmo~ and \lmo. 
For \nmo, the crystal structure down to 1.5~K was confirmed as a charge and Jahn-Teller orbital ordered state, as reported previously by room temperature single crystal x-ray diffraction measurements \cite{Awaka2005}. Above the magnetic transition temperature, diffuse scattering corresponding to a 2D short-range spin ordering is observed between 40 and 75~K. Below 40~K, the magnetic Bragg peaks, which are indexed by using a commensurate propagation vector $k_{\rm C} = (0.5, -0.5, 0.5)$, start to grow with decreasing temperature. Below 10~K, another set of magnetic Bragg peaks appears at incommensurate positions indexed with $k_{\rm IC} = (0, 0, 0.216)$. Based on the NPD data, we propose two AFM sublattices to explain the AFM ground state in \nmo. As for no-charge ordered \lmo, our neutron diffraction results reveal a lack of long-range magnetic ordering even at 1.5~K. 


\section{\label{sec:exp}Experimental methods}

\subsection{\label{sec:level2} Synthesis} 

\nmo~was prepared by a solid-state reaction under a high pressure, starting from \ce{Na2O2} and \ce{Mn2O3} powders \cite{Awaka2005}. A mixture of the two powders was packed in an Au-capsule in an Ar-filled glove-box, before being heated at 1300\doc~for 1 h under $P$ = 6~GPa, finally quenched to room temperature. For CFO-type \lmo, a spinel-type \lmo~was at first prepared by a conventional solid-state reaction using \ce{Li2CO3} and \ce{Mn2O3} powders \cite{Yamaura2006}. A mixture of precursors was heated at 850\doc~at ambient pressure. After the obtained spinel-type \lmo~powder was packed in an Au-capsule, heated at 1300\doc~for 1 h under $P$ = 6~GPa, and finally quenched to room temperature. 

\subsection{\label{sec:level2} Magnetic susceptibility}

Magnetization measurements as a function of temperature were performed with a superconducting quantum interference device (SQUID) magnetometer (MPMS, Quantum Design) in zero-field-cooled warming (zfcw) or field-cooled warming (fcw) or field-cooled cooling (fcc) modes in the temperature range between 5 and 300~K. 

\subsection{\label{sec:level2} Neutron powder diffraction}

Neutron powder diffraction patterns were recorded on the HRPT \cite{Fischer2000} and DMC diffractometers (Paul Scherrer Institute, Switzerland) with the wave length $\lambda$ = 1.886 and 1.494 \AA~ and $\lambda$ = 2.45 and 4.20 \AA, respectively, in the temperature range between 6 and 200~K. For both experiments the same sample batches as well as sealed $\diameter$6~mm vanadium sample cans were utilized. Data analysis and Rietveld refinements were performed with software tools from the FullProf Suite \cite{Rodriguez-Carvajal1993a}. Symmetry analysis was carried out using the Bilbao Crystallographic Server \cite{Aroyo2006, Kroumova2003} and ISODISTORT \cite{Campbell2006}.

\section{\label{sec:results}Results}

\subsection{\label{sec:crystal} Crystal structure of \nmo} 

\begin{figure*}[ht]
   \begin{center}
     \includegraphics[width=\linewidth]{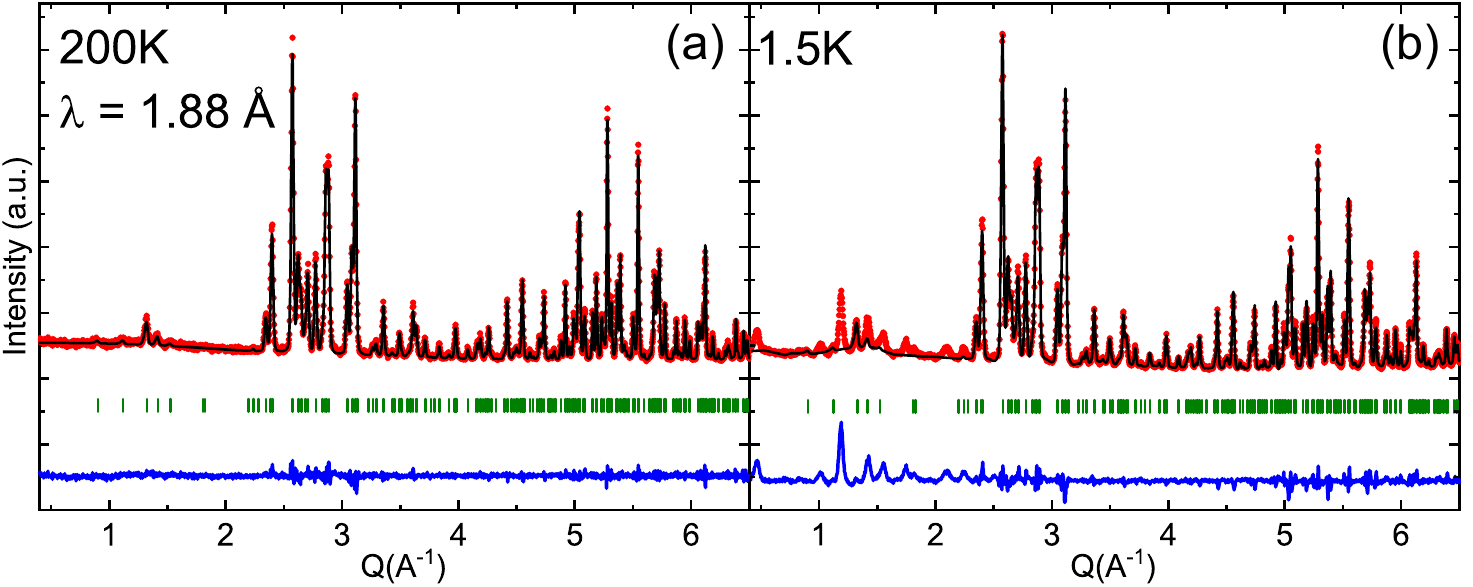}
   \end{center}
   \caption{Rietveld refinement of neutron diffraction data recorded on HRPT (PSI) of \nmo~at (a) $T=200$~K and (b) $T=1.5$~K using $\lambda$ = 1.88 \AA. Note that the magnetic peaks that are present in the 1.5~K diffraction pattern are here left unindexed.  Experimental data: open circles, calculated profile: continuous line, allowed Bragg reflections: vertical marks. The difference between the experimental and calculated profiles is displayed at the bottom of graph where a series of magnetic peaks are clearly visible.}
   \label{200K}
 \end{figure*}

\begin{figure*}[ht]
   \begin{center}
     \includegraphics[width=\linewidth]{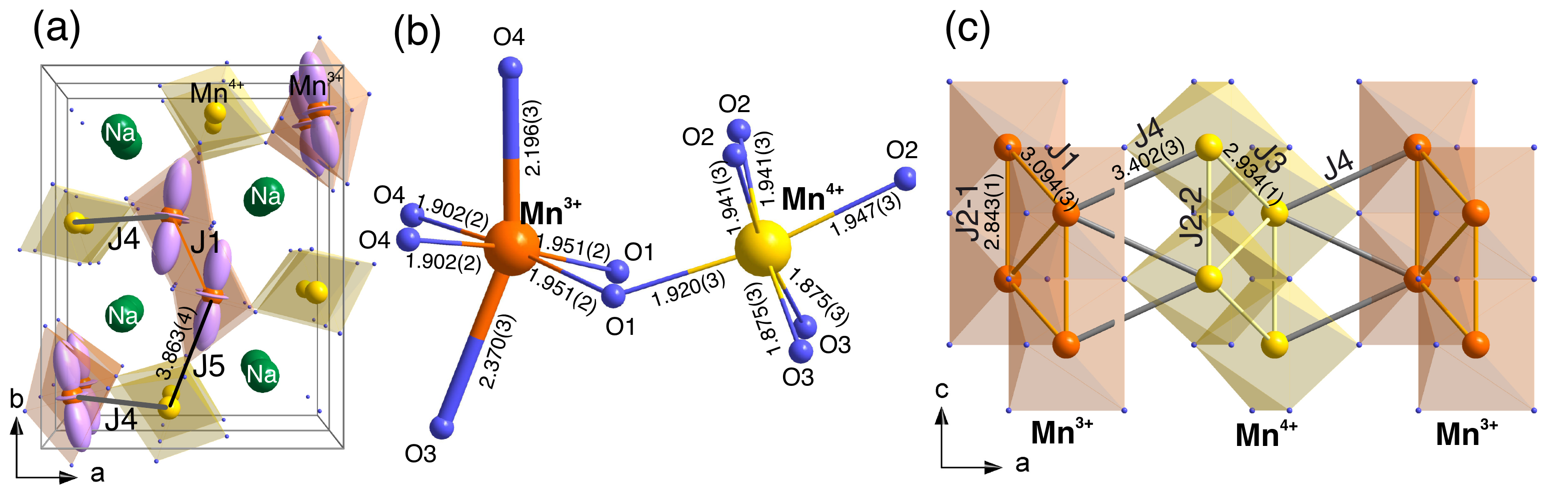}
   \end{center}
   \caption{Polyhedra representation of the CFO-type \nmo~structure projected (a-b) along [001] and (c) along [010]. The structure shows a charge ordering state of Mn$^{3+}$/Mn$^{4+}$ (orange and yellow polyhedra and atoms in figures, respectively), Na (green atoms), and the $d_{\rm z^2}$ orbital ordering (purple ellipsoid). In (b), the thick lines highlight the elongated Mn$^{3+}$-O distances. For clarity, only a part of a chain is shown in (c). The possible different magnetic exchanges paths, i.e. Mn-Mn bonds ($J_{\rm i}$) are shown by lines. The thick gray and black bonds show that the connections between Mn$^{3+}$ and Mn$^{4+}$ ladders corresponding to $J_4$ and $J_5$, respectively. The orange and yellow bonds links between same Mn sites. The labelled bond distances (in \AA) are from structural refinements at 200~K.}
     \label{structfig}
 \end{figure*}

The crystal structure of \nmo~at 200~K was refined in the orthorhombic \textit{Pnam} space group (No.62) with \ac~= 8.870 (1) \AA, \bc~= 11.227 (1) \AA and \cc~= 2.843 (1) \AA~starting from reported unit cell and atomic coordinates \cite{Awaka2005}. This structural model provides excellent fitting to the high-resolution NPD data recorded at HPRT [Fig. \ref{200K} and Table \ref{tablestructure}], and the corresponding crystal structure is illustrated in Fig. \ref{structfig}. This confirms that \nmo~is in a CFO-type structure with a charge ordering of Mn$^{3+}$ ($t_{2g}^3$ $e_g^1$) and Mn$^{4+}$ ($t_{2g}^3$ $e_g^0$). 
In the CFO structure, two different \ce{Mn2O4} zigzag chains are formed by a network of edge-sharing \ce{MnO6} octahedra (Mn-Mn bond labelled as $J_2$) aligned along the \cc-axis [Fig. \ref{structfig}(a, c)]. These four zigzag chains are linked by corner-sharing octahedral (to connect equivalent Mn site: next nearest-neighbor Mn1 ($J_1$) and Mn2 ($J_3$), non-equivalent Mn site: $J_4$ and $J_5$ depending on the direction) to form a 3D framework structure and make an irregular hexagonal 1D tunnel along the \cc-axis, in which \ce{Na^+} ions are located. 

The Mn$^{3+}$ and Mn$^{4+}$ ions are fully ordered (within the refinement standard deviation), in contrast to the other CFO-type \ce{Na}$B^{3+}B^{4+}$\ce{O4}  ($B$ = V, Ti, Cr) materials. 
As a results, in the \nmo~lattice, two crystallographically different Mn sites are clearly distinguishable, which are coordinated by six oxygen atoms [Fig. \ref{structfig}(b)]. 
As already reported in \cite{Awaka2005}, the \ce{Mn$^{3+}$O6} octahedron (manganese labelled as Mn1) is strongly elongated due to stabilization of the $d_{\rm z^2}$ orbital, i.e. Jahn-Teller distortion of the Mn$^{3+}$ ions. 
That is, among the six Mn1-O bonds of the Mn$^{3+}$\ce{O6} octahedron, four Mn1-O distances ranges between 1.90 \AA~and 1.95 \AA~and two Mn1-O distances are larger than 2.15 \AA, 
resulting in an average distance of 2.05 \AA~at 200~K, as summarized in Table \ref{tablelength}.  
The stabilized $d_{\rm z^2}$ orbitals are aligned in the $ab$-plan and form a zig-zag pattern on the [001] projection, consisting of alternating ladder rows with two different orbital orientations [Fig. \ref{structfig}(a)]. 
Within the ladders, the distorted triangular lattice is formed by the Mn$^{3+}$ ions as illustrated in Fig. \ref{structfig}(c). 
More correctly, three edges of each triangular lattice correspond to one short Mn1-Mn1 distances of 2.84 \AA ($J_{\rm 2-1}$) and two longer Mn1-Mn1 distances of 3.09 \AA ($J_{\rm 1}$). 
In this edge-sharing octahedra network, the half-filled $t_{\rm 2g}$ orbitals are directed along the short Mn-Mn distance in the edge-sharing octahedra ($J_{\rm 2-1}$), 
and should favour strong antiferromagnetic direct exchange interaction, rather than a superexchange interaction through the Mn-O-Mn pathway (the angle of $\sim$ 90\dd) \cite{Goodenough1963}.  
 
In contrast, the \ce{Mn$^{4+}$O6} octahedron (labelled as Mn2) poses very small distortion with the average distance between Mn2 and O of 1.92 \AA. 
In order to compare the distortion of an octahedron quantitatively, one could use the distortion parameter:

\begin{eqnarray}
\Delta d = 1/6\sum_{n=1,6} \left[(d_n - \langle d \rangle)/{\langle d \rangle}\right]^2
\label{eq:distortion}
\end{eqnarray}

which corresponds to the deviation of Mn-O distances with respect to the average distance. In fact, $\Delta d$ of \ce{Mn$^{3+}$O6} is 74 $\times$ 10$^{-4}$, whereas for \ce{Mn$^{4+}$O6}, $\Delta d$  is 2.5 $\times$ 10$^{-4}$, i.e. 1/30. Note than $\Delta d$ for many Jahn-Teller distorted manganites shows $\Delta d$ $>$ 30 $\times$ 10$^{-4}$ \cite{Matsubara2017a, PhysRevB.57.R3189}. The triangular ladder of Mn$^{4+}$ is also much more regular (Mn2-Mn2 distances: 2.8 - 2.9 \AA) than that of \ce{Mn$^{3+}$} (Mn1-Mn1 distances: 2.8 - 3.1 \AA).

There is no clear evidence for a structural transformation down to 1.5~K, although additional diffraction peaks appear below \TN~due to a magnetic transition, as discussed later. Figure~\ref{200K}(b) shows the refinement result for the diffraction pattern recorded at 1.5~K. 
The reliability of fitting the data at 1.5~K is slightly worse than that at 200~K, 
because the reduced $\chi^{2}_{\rm fit}$ for the data at 1.5~K is 5.34 and 2.17 at 200~K. 
The quality of fit is slightly improved by including a monoclinic distortion ($\beta$ = 90.1(1)\dd) (using $\Gamma$ point subgroup: $P2_1/c$ or $P2_1/m$). 
However, the obtained monoclinic distortion is too small to clarify with the resolution of the present NPD data. 
Therefore, the low-temperature crystal structure is considered to be the same as the one at 200~K ($Pnam$). 
This is also supported by the fact that the data of Tables \ref{tablestructure} and \ref{tablelength}, the unit cell volume is found to contract only by 0.40 (1) $\%$, 
when temperature is decreased from 200 to 1.5~K. This means that both the charge ordering and orbital pattern remains unchanged at the onset of magnetic order. 
A possible structural transition induced by magnetic ordering will be discussed in the next section.

\begin{table}[hb]
\caption{\label{tablestructure} Structural parameters of \nmo~at 200 and
1.5~K from Reitveld refinements of HRPT data. The space group is \Pnam~and all the atoms are in site $4c~ (x, y, 1/4)$.}
\begin{ruledtabular}
\begin{tabular}{ccdd}
&&
\multicolumn{1}{c}{\textrm{200K}}&
\multicolumn{1}{c}{\textrm{1.5K}}\\
\hline
& \ac~(\AA) &  8.870 (1) & 8.861 (1) \\
& \bc~(\AA) & 11.227 (1) & 11.197 (1) \\
& \cc~(\AA) & 2.843 (1) & 2.841 (1) \\
& \Vc~(\AA$^3$) & 283.1 (1) & 281.9 (1) \\
Na& \x  & 0.2473 (4)  & 0.2472 (4) \\
& \y  & 0.3345 (3)  & 0.3352 (3) \\
& \textit{B} (\AA$^2$)  & 0.52 (6)  & 0.33 (6)\\
Mn1& \x  & 0.0714 (3)  & 0.0716 (4) \\
& \y  & 0.1086 (3)  & 0.1088 (3) \\
& \textit{B} (\AA$^2$)  & 0.01 (6)  & 0.01 (6) \\
Mn2& \x  & 0.0806 (4)  & 0.0807 (4) \\
& \y  & 0.5949 (3)  & 0.5956 (3) \\
& \textit{B} (\AA$^2$)  & 0.01 (6)  & 0.14 (7)\\
O1& \x  & 0.2864 (2)  & 0.2870 (2) \\
& \y  & 0.6478 (2)  & 0.6478 (2) \\
& \textit{B} (\AA$^2$)  & 0.09 (4)  & 0.08 (4)\\
O2& \x  & 0.3832 (2)  & 0.3849 (3)\\
& \y  & 0.9809 (2)  & 0.9802 (2)\\
& \textit{B} (\AA$^2$)  & 0.09 (4)  & 0.21 (4)\\
O3& \x  & 0.4747 (2)  & 0.4745 (2)\\
& \y  & 0.1947 (2)  & 0.1942 (2) \\
& \textit{B} (\AA$^2$)  & 0.30 (4)  & 0.35 (4)\\
O4& \x  & 0.0683 (2)  & 0.0689 (3)\\
& \y  & 0.9132 (2)  & 0.9135 (2)\\
& \textit{B} (\AA$^2$)  & 0.16 (4)  & 0.29 (4)\\
& \textit{R}$_{\rm Bragg}$ ($\%$)   & 4.19  & 4.69 \\
& $\chi^{2}_{\rm fit}$    & 2.17  & 5.34\\

\end{tabular}
\end{ruledtabular}
\end{table}

\begin{table}[hb]
\caption{\label{tablelength} Selected distances (\AA) and angles (\dd) in \nmo~at 200 and 1.5~K. The distortion parameter is calculated for each manganese octahedron as $\Delta d = 1/6\sum_{n=1,6} \left[(d_n - \langle d \rangle)/{\langle d \rangle}\right]^2$, with ${\langle d \rangle}$ being the average distance between Mn and O.}
\begin{ruledtabular}
\begin{tabular}{ccdd}
& \multicolumn{1}{c}{\textrm{200~K}}&
\multicolumn{1}{c}{\textrm{1.5~K}}\\
\hline
Mn1 - O1       & 1.951(2)   & 1.950(4)    \\
Mn1 - O3       & 2.370(3)   & 2.372(6)     \\
Mn1 - O4       & 2.196(3)   & 2.188(6)    \\
Mn1 - O4       & 1.902(2)   & 1.909(4)     \\
average        & 2.046(1)   & 2.046(2)    \\
$\Delta d$ $\times$ 10$^{-4}$    & 74.4  & 72.6  \\
Mn2 - O1       & 1.920(3)   & 1.911(6)    \\
Mn2 - O2       & 1.947(3)   & 1.952(6)     \\
Mn2 - O2       & 1.941(3)   & 1.956(4)     \\
Mn2 - O3       & 1.875(2)   & 1.853(4)     \\
average        & 1.916(1)   & 1.914(2)    \\
$\Delta d$ $\times$ 10$^{-4}$        & 2.537  & 5.723   \\
Mn-Mn along    &            &             \\
c ($J_{\rm 2-1}$, $J_{\rm 2-2}$) & 2.843 (1)  & 2.841 (1)   \\
Mn1 - Mn1 ($J_1$) & 3.094 (3)  & 3.088(6)     \\
Mn1 - O1 - Mn1 & 93.54 (9)  & 93.48(17)    \\
Mn1 - O4 - Mn1 & 96.70(9)   & 96.17(18)    \\
Mn2 - Mn2 ($J_3$) & 2.934(4)   & 2.957(6)     \\
Mn2 - O2 - Mn2 & 94.19(11)  & 93.12(16)    \\
Mn2 - O3 - Mn2 & 98.62(10)  & 100.12(16)   \\
Mn1 - Mn2 ($J_4$) & 3.402(3)   & 3.402(7)    \\
Mn1 - Mn2 ($J_5$) & 3.863(4)   & 3.833(6)     \\
Mn1 - O1 - Mn2 & 123.0(2)   & 123.5(4)    \\
Mn1 - O3 - Mn2 & 130.7(2)   & 129.9(4)   
\end{tabular}
\end{ruledtabular}
\end{table}

\subsection{\label{sec:diffuse} Magnetic order of \nmo: Diffuse scattering}

  \begin{figure}[t]
   \begin{center}
     \includegraphics[width=\hsize]{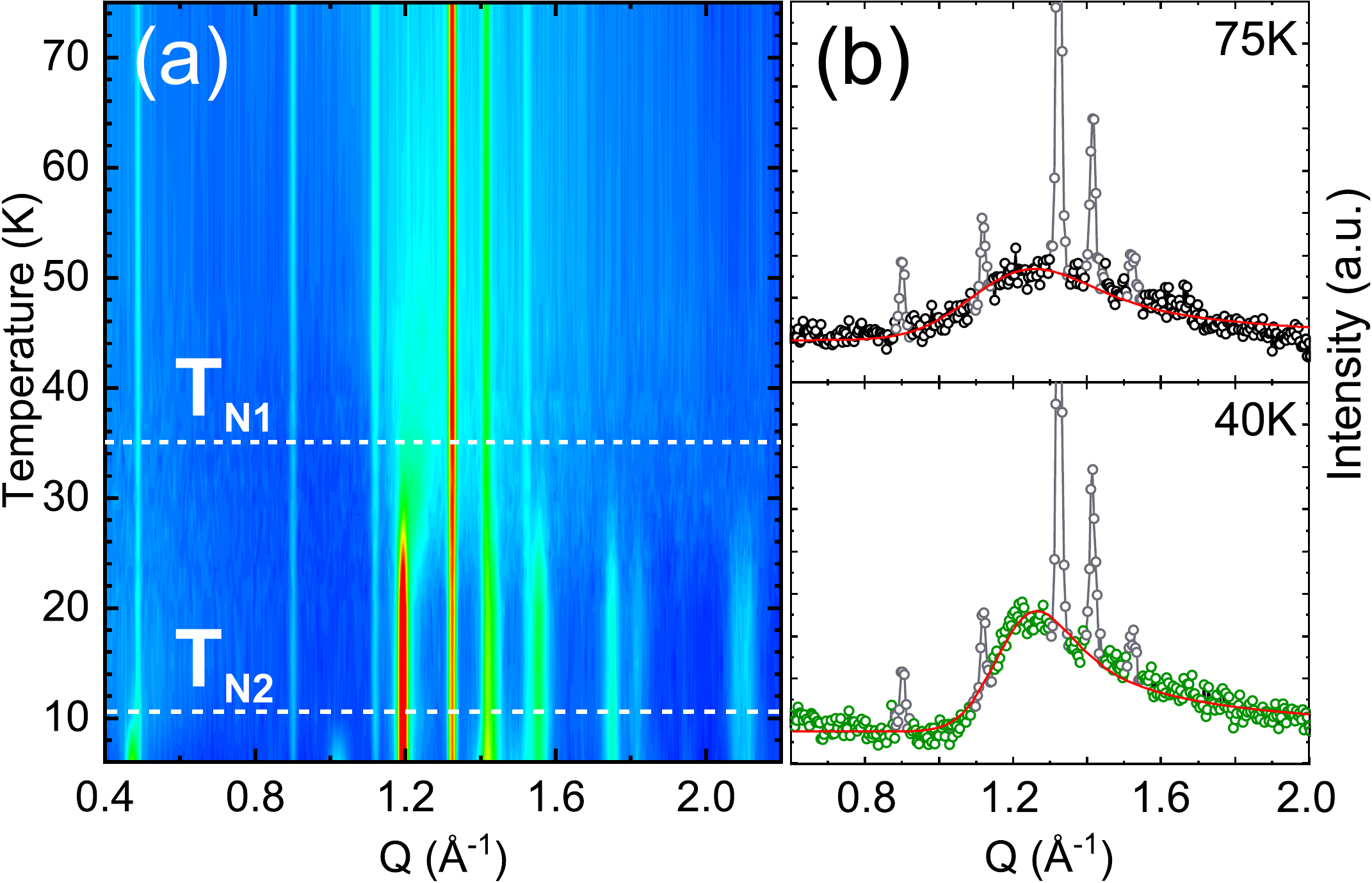}
   \end{center}
   \caption{(a) Temperature evolution of the powder neutron diffraction patterns of \nmo~(DMC $\lambda$ = 2.45 \AA) between 75 and 6~K in the 0.4 - 2.2 \AA$^{-1}$ range.
    \TTN~$\approx35$~K and \TTTN~$\approx11$~K are marked by the horizontal white dashed lines. 
    (b) Neutron diffraction patterns at two temperatures above \TTN~$\approx35$~K in \nmo. 
    Circles show experimental data, Bragg peaks corresponding to the crystal structure are masked in gray and the solid red line is the fitted Warren line shape given by \cite{Warren1941}}
   \label{colormap}
 \end{figure}

  \begin{figure}[ht]
   \begin{center}
     \includegraphics[width=\hsize]{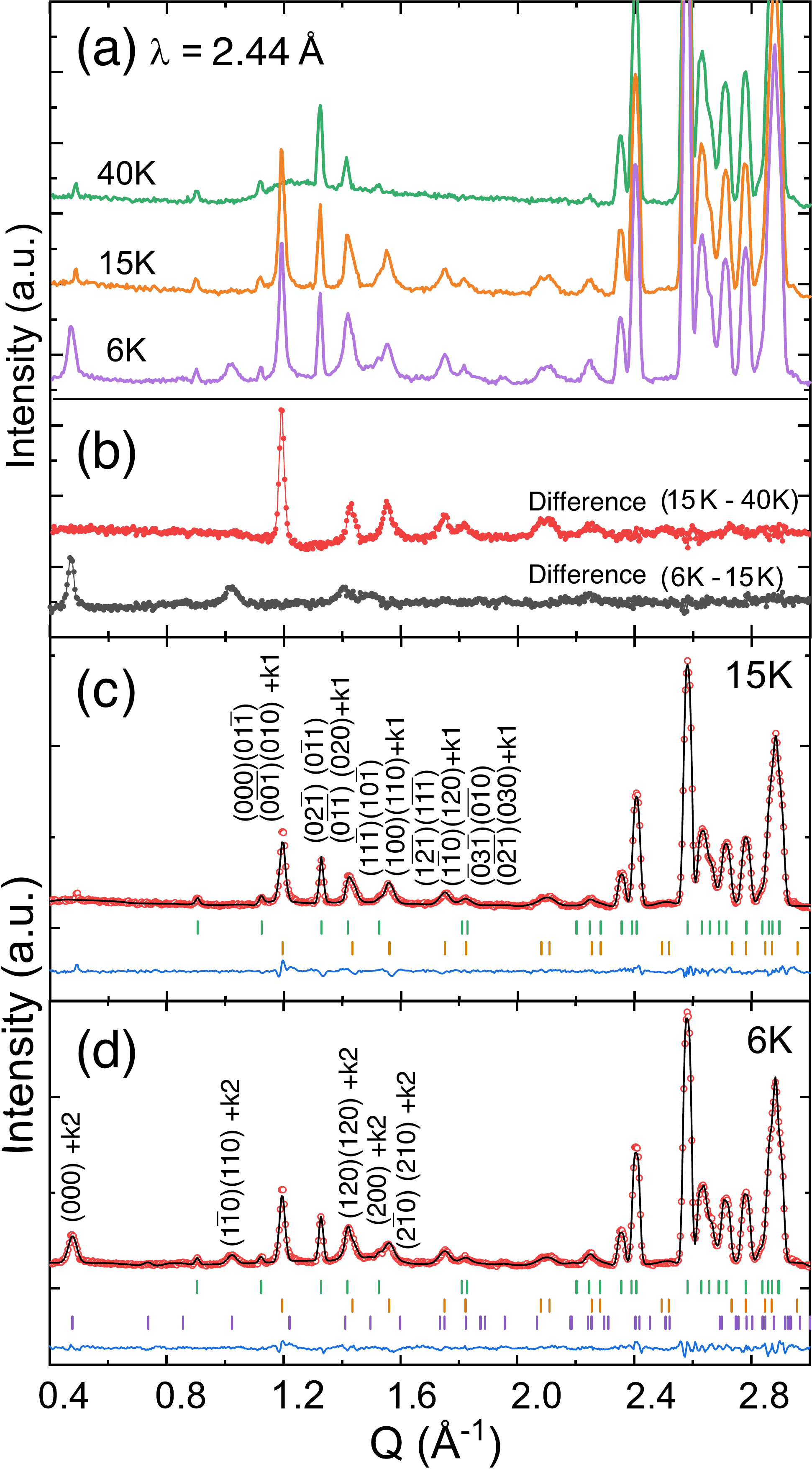}
   \end{center}
   \caption{(a) Neutron powder diffraction patterns recorded at 6, 15 and 40~K, and (b) their differences (DMC $\lambda$ = 2.45 \AA). Rietveld refinement of the neutron diffraction data of \nmo~at (c) 15~K and (d) 6~K. Green, orange and purple ticks indicate Bragg peak positions of crystal, commensurate magnetic and incommensurate magnetic structure, respectively.}
   \label{diffuse}
 \end{figure}

In order for further understanding the magnetic transition of \nmo~\cite{Sugiyama2009b}, 
we have performed NPD as a function of temperature between 6 and 200~K on DMC in PSI [Fig.~\ref{colormap}(a)]. 
While there are no magnetic diffraction peaks above 40~K, the absence of three-dimensional long-range magnetic order is confirmed, a diffuse scattering is clearly observed, indicating the formation of short-range ordering. 
The shape of such magnetic diffuse scattering is asymmetric with a maximum at $Q$ $\sim$ 1.2 \AA$^{-1}$ and is well fitted with a Warren function \cite{Warren1941} [see Fig.~\ref{colormap}(b)]. 
This is characteristic for a 2D short-range order. In fact, the fit provides that, as temperature decreases from 75 to 40~K, the 2D correlation length increases from $\sim$ 18 \AA~to $\sim$ 28 \AA. 
Such correlation length is comparable to those for the other related compounds with a similar diffraction profile, such as \ce{NaMnO2} \cite{Giot2007}, \ce{CuMn2O4} \cite{Terada2011}, 
which has a layered transition metal triangular lattice structure, and \ce{Dy2TiO5} \cite{Shamblin2016} with Dy triangular ladder configuration. 
Moreover, the previous \musr~studies of \nmo~\cite{Sugiyama2009b} revealed the presence of a disordered state at temperatures between \TN~and $\sim$ 60~K, 
which is fully consistent with the current NPD results. 
More detailed interpretation of this diffuse scattering is mentioned below in Sec.~\ref{magorder}.

%
%


\subsection{\label{magorder} Magnetic order of \nmo: Magnetic structure refinement}

As temperature decreases from 40~K, the magnetic diffuse scattering is gradually transformed into magnetic Bragg peaks [see Fig.~\ref{colormap}(a)], 
being consistent with the previous \musr~result. Furthermore, there are clearly two magnetic transitions, \TTN~= 35~K and  \TTTN~= 11~K. 
Noteworthy the $\chi$(T) curve shows the cusp of antiferromagnetic transition around $\sim$ 15~K, which is much lower than first \TN~determined with \musr~and NPD. 
Indeed, as seen in Fig.~\ref{sus}(b), not only \TTTN~is well evidenced in the differential susceptibility [$d\chi$/$dT$]($T$) curves around \TTTN ($\sim$ 9K), 
but also small anomaly is observed around \TTN ($\sim$ 40K) in both zfc and fc modes (only zfc is shown). This is in agreement with the observation of the diffuse scattering by NPD.
Such vague anomaly observed around \TTN~is rather abnormal, considering that clear magnetic Bragg peaks observed by NPD. This might be related with the crystallinity (grain size) of the sample as reported on Ref.\cite{Matsubara2019a}.     

Figure~\ref{diffuse}(a) compares the diffraction patterns recorded at 40, 15 and 6~K in order to clarify the change in Bragg peaks across \TTN~and \TTTN. 
The difference of diffraction patterns between 40 and 15~K [Fig.~\ref{diffuse}(b)] clearly demonstrates the appearance of a set of magnetic Bragg peaks, 
which are all indexed by a commensurate (C) propagation vector $k_{\rm C}$ = (0.5, -0.5, 0.5). 
On the other hand, the difference of patterns between 15 and 6~K is indexed using an incommensurate (IC) propagation vector $k_{\rm IC}$ = (0, 0, 0.216 (1)). 
Surprisingly, despite the magnetic transition from the C-AF phase to the IC-AF phase at \TTTN, the C-AF diffraction pattern does not change even below \TTTN~[see details in Fig.~\ref{diffuse}(b)]. 
This means that the C-AF structure is preserved even in the IC-AF ordered state appeared below \TTTN. 
This also implies that these two magnetic sublattices are mutually independent.

\begin{figure*}[ht]
  \begin{center}
     \includegraphics[width=\textwidth]{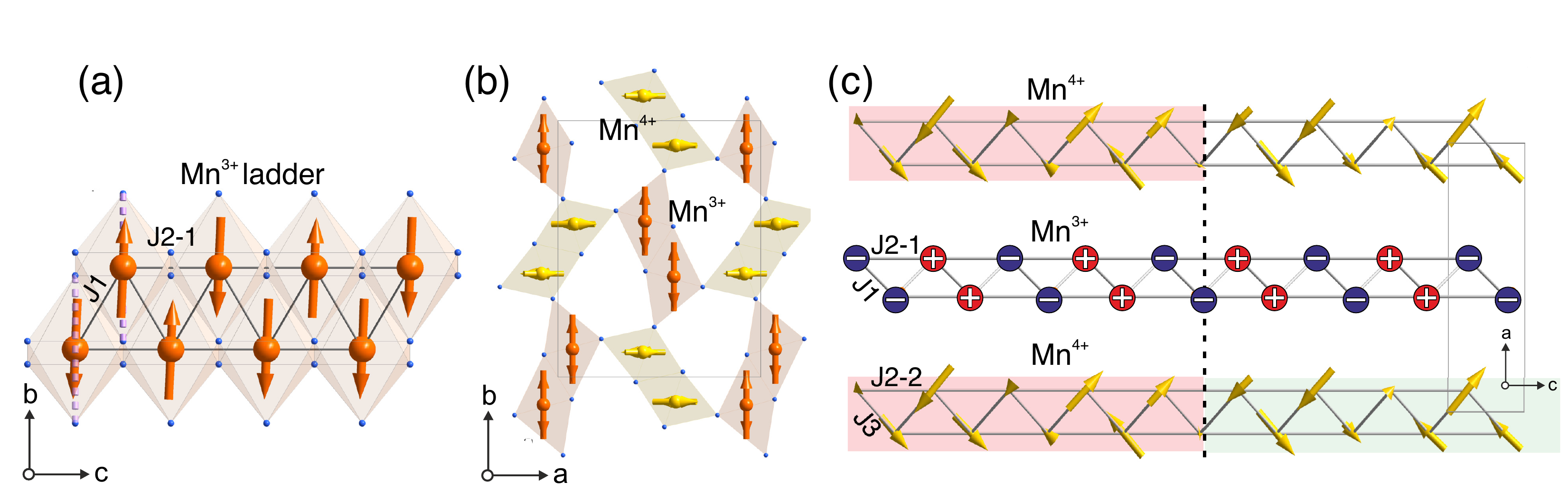}
   \end{center}
   \caption{Magnetic structure of \nmo~at 6~K. (a) $k_{\rm C}$ commensurate spin configuration along Mn$^{3+}$ ladder. The purple dash line is the guide to indicate the direction of $z$-orbital. (b) Overall spin structure of \nmo~along \cc-axis in the honeycomb mesh. 
   (c) Two Mn ladders along \cc-axis. The dashed line highlights the spin inversion plane of $k_{\rm IC}$ incommensurate spin arrangement. The magnetic moments of commensurate spins along \bc~direction are symbolized by + or − signs depending on their spin direction.}
   \label{magstructure}
 \end{figure*}

To constrain the number of solutions for the magnetic models, symmetry analysis was performed using $k_{\rm C}$ and $k_{\rm IC}$, for each Wyckoff site $4e$ of Mn. 
The commensurate $k_{\rm C}$ ordering is described as antiferromagnetic chains running along the \cc-axis ($J_{\rm 2}$). 
Next-neighbor Mn arrange antiferromagnetically along [01-1] and ferromagnetically along [011] ($J_{\rm 1}$). 
In other words, the Mn spin chains makes -[AFM-FM-AFM-FM]- zigzag pattern along the \cc-axis [see Fig.~\ref{magstructure}].

\begin{figure}[ht]
  \begin{center} 
     \includegraphics[width=\hsize]{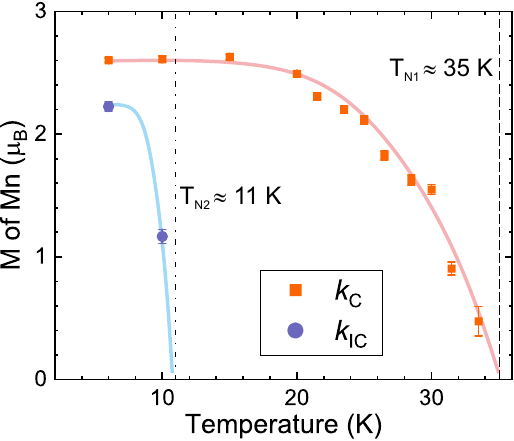}
   \end{center}
   \caption{Temperature evolution of the Mn$^{3+}$ and Mn$^{4+}$ ordered magnetic moment (from Rietveld refinement results). Solid lines are guides to the eye, while the dashed line and dash-dotted line are at \TTN = 35~K and \TTTN = 11~K, respectively.}
   \label{orderpara}
 \end{figure}

The proposed model is expressed by the $C_ac$ Shubnikov group (BNS 9.41). 
Using this spin arrangement, the diffraction pattern recorded at 15~K is well fitted with only one Mn site [Fig. \ref{diffuse}(c)], 
leading to the best $R_{\rm Bragg}=1.9$\% and $R_{\rm mag}=8.4$\%. 
Such high $R_{\rm mag}$ is due to the trace of diffuse scattering on the shoulder of several magnetic Bragg peaks, especially for the one at $Q$ $\sim$ 1.2 \AA$^{-1}$. 
At 15~K, the Rietveld refinement [Fig.~\ref{diffuse}(c)] yields 
$m_{\rm y}=2.6(1)$ and $m_{\rm z}=0.4(1)$, which gives a total ordered magnetic moment ($\mu^{\rm C}_{\rm ord}$ =) 2.63(2)~$\mu_{\rm B}$. 
Note that $m_{\rm x}$ is set to zero even it is not restricted by symmetry, because refining this component does not improve the fitting of our NPD data. 
On the other hands, the small component $m_{\rm z}$ is necessary to obtain better fit for the magnetic Bragg peaks.
The size of $\mu^{\rm C}_{\rm ord}$ is unchanged down to 6~K [Fig.~\ref{orderpara}]. The main component of $\mu^C_{\rm ord}$, i.e. $m_{\rm y}$ in the two adjacent chains follows approximately the orbital pattern of Mn$^{3+}$ as described earlier. This spin direction is understandable by considering the single ion anisotropy of Mn$^{3+}$ in the elongated octahedra along the local \textit{z} axes due to the Jahn-Teller distortions.
Besides the spin is slightly off ($\sim$ 14 \dd) from the local \textit{z} orbital. This deviation possibly results from the competition among magnetic anisotropy, frustration, and 90\dd~superexchange interaction.

For the incommensurate $k_{\rm IC}$ order, there are four one-dimensional irreducible representations (Irreps) and three basis functions for each Irreps. 
Only the symmetry modes spanned by the Irreps $\Gamma_4$ provided good agreement with the experimental data. 
Table~\ref{irreps} lists the basis vectors of Irreps $\Gamma_4$, obtained by the projection operators method. 
The IC-AF order was firstly fitted using only one Mn site following upon the C-AF phase results. 
Figure~\ref{magstructure} shows the spin structure that gives the best agreement factor $R_{\rm mag}$ = 13.6\% with the experimental data [see Fig.~\ref{diffuse}]. This model eventually corresponds to a spin density wave (SDW), obtained by mixing the basis function $\psi_1$ and $\psi_3$ of real components only. 
At 6~K, $m_{\rm x}=1.4(1)$ and $m_{\rm z}=1.5(1)$ are obtained by a refinement of the magnetic structure [Fig.~\ref{diffuse}(d)]. 
For this model, the wave maximum amplitude is about 2.0(1) $\mu_{\rm B}$ (= $\mu^{\rm IC}_{\rm ord}$) and $\mu^{\rm IC}_{\rm ord}$ lay in the \textit{ac}-plane with the spin modulation along [101] and [10-1] directions. 
The periodicity of the IC modulation is $\sim$ 4.63 $\times$ \cc~and the intraleg interaction is AFM with the domain boundary shown as the dotted line in Fig.~\ref{magstructure}(c). 
The incommensurate spin structure is also described involving $\psi_2$ (corresponding to the $m_{\rm y}$ component) based on the symmetry analysis. 
An unconstrained refinement using DMC data ($\lambda=2.45$~\AA), using only the symmetry restrictions of $\Gamma_4$, showed that the $m_{\rm y}$ component of $\mu^{\rm IC}_{\rm ord}$ is nearly zero. 
As a result, in the final refinement stage, the $m_{\rm y}$ component was set to zero.  

Because of the fact that $\mu^{\rm C}_{\rm ord}$ is greater than $\mu^{\rm IC}_{\rm ord}$, 
the $k_{\rm C}$ magnetic structure is assumed to be of Mn$^{3+}$ (Mn1) site while the $k_{\rm IC}$ phase for Mn$^{4+}$ (Mn2). Both magnetic moments are smaller than the expected values for the fully ordered state, which is coherent with remnant diffuse scattering. 
Such discrepancy is also reported for \textit{e.g.} \ce{NaMnO2} (2.92~$\mu_{\rm B}$ for Mn$^{3+}$) \cite{Giot2007}. 
It is worthwhile to mention the possibility of slight mixing between Mn$^{3+}$ and Mn$^{4+}$ (both Mn sites can form C-AF and IC-AF configuration). 
This is well-known issue in charge ordered manganites and such mixing often leads to smaller $\mu_{\rm ord}$ than the expected value; 
\textit{e.g.} \ce{NaMn7O12} (2.85~$\mu_{\rm B}$ for Mn$^{3+}$ and 2.40~$\mu_{\rm B}$ for Mn$^{4+}$ at 10~K) \cite{Prodi2004}.  

\begin{table*}
\caption{\label{irreps} Basis functions for axial vectors associated with the irreducible representation $\Gamma_4$ for Wyckoff site 4$c$ and $k_{\rm IC} = (0, 0, 0.216)$. $ a = \mathrm{e}^{+i\pi q}$} 
\begin{ruledtabular}
\begin{tabular}{ccccc}
$\Gamma_4$ & Mn2-1 & Mn2-2 & Mn2-3 & Mn2-4 \\
\hline
& ($x y z$) & ($-x+1,-y+1, z+\dfrac{1}{2}$) & ($-x+\dfrac{1}{2},y-\dfrac{1}{2},z+\dfrac{1}{2}$) &  ($x+\dfrac{1}{2},-y+\dfrac{3}{2},z$)  \\
\hline
$\psi_1$& ($100$) & ($a^*00$) & ($a^*00$) & (100)  \\
$\psi_2$& (010) & ($0a^*0$) & ($0-a^*0$) & (0-10)\\
$\psi_3$& (001) & ($00-a^*$) & ($00-a^*$) & (001)\\
\end{tabular}
\end{ruledtabular}
\end{table*}      

Among the asymmetric diffuse scattering around the commensurate magnetic peaks at temperature above \TTN, 
the strongest one was observed at $Q$ $\sim$ 1.2 \AA$^{-1}$ ((000)(01-1)(00-1)(010)+$k_{\rm C}$)) (see Sec.~\ref{sec:diffuse}). 
This position suggests the short-range correlations start generating along the chain axis ($bc$-plane), compatible with the magnetic interaction along this direction ($J_{2-1}$: corresponding to the commensurate ordering). 


The feature of the $k_{\rm C}$ magnetic structure is described in Fig.~\ref{magstructure}, with the antiferromagnetic chains parallel along the \cc-axis. 
All the intraleg Mn-O-Mn supercharge interactions are rather weak, because the angel of the Mn-O-Mn bonds is nearly 90\dd. 
However, for the case of the nearest-neighbour interaction ($J_2$), 
both 90\dd~ Mn1-O1-Mn1  and Mn1-O4-Mn1 bonds are formed by a direct overlap of the $d_{\rm xy}$ (or $d_{\rm yz}$ or $d_{\rm zx}$) a strong AFM direct-exchange interactions are possible \cite{Goodenough1963}, 
which causes a geometrical frustration configuration. 
Due to such frustration on the triangular lattice in the two-leg ladder, 
a magnetoelastic coupling is expected to appear below \TTN. 
The $C_{a}c$ Shubnikov group has a monoclinic supercell $a_{\rm m}\sim5.68$~\AA, $b_{\rm m}\sim17.7$~\AA, $c_{\rm m}\sim11.5$~\AA~ and $\beta_{\rm m}\sim104$\dd. 
In this supercell, there is a $\Gamma^{3+}$ displacement mode (with $k$-vector (0,0,0) and direction (a) corresponding to the isotropy subgroup $P2_1/c$ (No. 14)) 
which allowed a shift of Mn atoms along \cc-direction, but two Mn rows (within ladders) shift in the opposite $z$ directions. 
This distortion component could deform the triangular lattices in order to stabilize the $J_1$ magnetic interaction. 
However, as mentioned in Sec.~\ref{sec:crystal}, the monoclinic distortion in the crystal structure is not evident in the present NPD data. 
Further studies at low temperatures using single crystal synchrotron X-ray and electron diffraction techniques will be valuable for investigating the magnetoelastic coupling of this compound.     



For the $k_{\rm IC}$ magnetic structure, although the coupling within the Mn$^{4+}$ ladder is rather complicated to interpret, one could compare with a mixed valance CFO-\ce{NaV2O4}, which poses a similar SDW magnetic structure with the IC propagation vector, $k = (0, 0.19, 0)$ \cite{Nozaki2010}. 
The direction of the ordered spins was found to be perpendicular to the ladder direction with the SDW modulation along the ladder. 
This configuration and modulation are similar to the current  Mn$^{4+}$ ladder case, except that \nmo~has magnetic components (roughly $\sim$ 30\dd) along the ladder direction. 
This might be due to the additional superexchange interaction (intermediate angle $\sim$ 130\dd) between Mn$^{3+}$ and  Mn$^{4+}$ ladders. In order to understand the detail of such spin coupling, further magnetic characterization will be required, such as single crystal studies, inelastic neutron scattering along with theoretical calculations and computer simulations. 
Finally, it could be also very interesting to perform 
additional investigations under high hydrostatic pressure to know 
how subtle externally induced structural distortions affect the magnetic correlations \cite{Forslund2019} in this complex material.

\subsection{\label{sec:level2} Neutron diffraction study of \lmo} 

\begin{figure}[ht]
  \begin{center} 
     \includegraphics[width=\hsize]{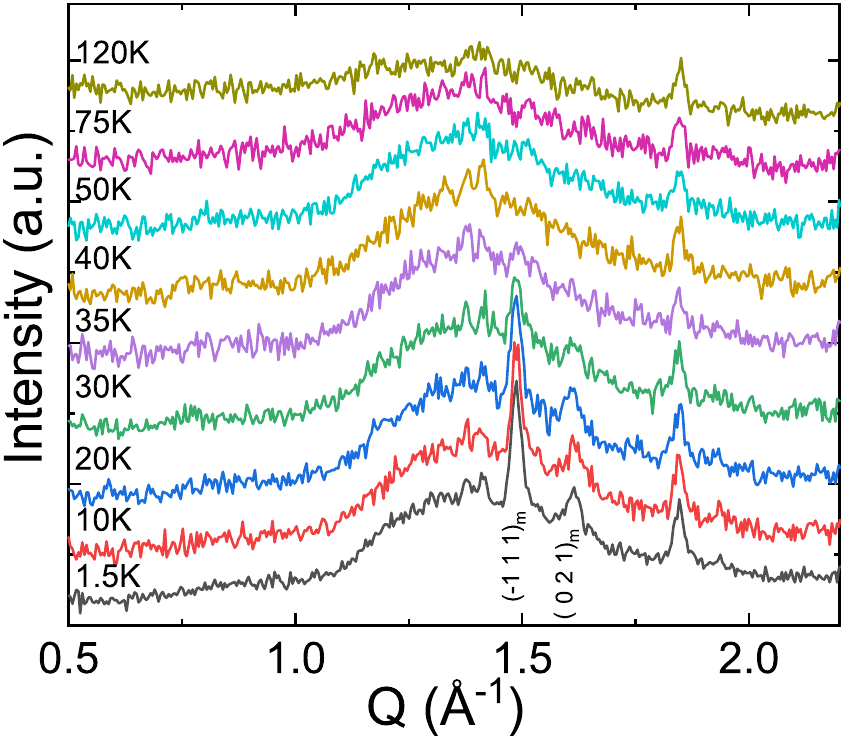}
   \end{center}
   \caption{Temperature evolution of the powder neutron diffraction patterns of \lmo~(DMC $\lambda$ = 2.45 \AA) between 1.5 and 120~K in the 0.5 - 2.2 \AA $^{-1}$ range. Two magnetic Bragg peaks of the \ce{Li2MnO3} impurity phase are indexed.}
   \label{LMO}
 \end{figure}

We have also performed a NPD study on the closely related CFO-type compound, \lmo in order to compare the magnetic nature within this family. 
The crystal structure of \lmo~was confirmed as a reported one \cite{Akimoto2009}. 
Unlike the \nmo, \lmo lacks clear charge and/or orbital order; 
thus both Mn$^{3+}$ and Mn$^{4+}$ are equally and randomly distributed on the two Mn sites. 
The resolution and $Q$-range of the present neutron diffraction study recorded on the DMC diffractometer ($\lambda$ = 2.45 \AA) are not enough to perform a proper structural analysis.  
Therefore, the Rietveld refinement was performed using reference structural parameters \cite{Mukai2019} yielding a reasonable fit result. 
The temperature dependent data clarified the absence of changes in the crystal structure at temperatures between 120 and 1.5~K. 
Moreover, the NPD result revealed the presence of the \ce{Li2MnO3} phase with the volume of 16.1$\%$ as a second phase. 
This is in good agreement with the previously reported XRD and \musr~results \cite{Mukai2019, Sugiyama2009b}. 
From the previous \musr~measurement of \lmo, the observed magnetic signal was in fact found to come from the Li$_2$MnO$_3$ phase \cite{Sugiyama2009b}. 
The present NPD results are fully consistence with the reported \musr~data; 
in other words, only magnetic Bragg peaks corresponding to the \ce{Li2MnO3} phase were observed below \TN~=~35~K \cite{Strobel1988}. 
Two main magnetic Bragg peaks of \ce{Li2MnO3} are indexed and shown in Fig.~\ref{LMO}. 
There is also clear diffuse scattering (maximum around $Q$ $\sim$ 1.3 \AA $^{-1}$) correlated with a background reduction, 
which indicate the appearance of short-range magnetic order. 
Below 50~K, such diffuse scattering remains unchanged down to 1.5~K, i.e. even below the magnetic transition temperature of \ce{Li2MnO3}. 
This suggests that the diffuse scattering is not originating from \ce{Li2MnO3} but an intrinsic feature of \lmo, 
which has a magnetic transition at \TN~=~44~K according to the \musr~results \cite{Sugiyama2009b}. 
However, further studies are required to get information on the detailed magnetic nature of \lmo. 

\section{\label{sec:level2} Discussion} 

\nmo~is found to exhibit two successive antiferromagnetic transitions at \TTN~= 39~K and \TTTN~= 11~K. 
The former is a transition to a C long-range Mn$^{3+}$ spin ordered state and the latter is a transition to an IC Mn$^{4+}$ spin ordered state. 
Our study indicates that these two Mn$^{3+}$/Mn$^{4+}$ magnetic lattices are independent each other. 
Back to inverse susceptibility (1/$\chi$) shown in Fig.~\ref{sus}(c), 
although the 1/$\chi$(T) curve is linear with temperature above 200~K, it deviates from such linear relationship below around 150~K, 
which is relatively high compared with the observed cusp (around 15~K). 
If the temperature evolution of the Mn moment in the Mn$^{3+}$ sublattice is independent of that in the Mn$^{4+}$ sublattice, 
each magnetization follows a Curie-Weiss behavior with its own Curie constant ($C$) and Weiss temperature ($\Theta$). 
Since $\Theta$ of the Mn$^{3+}$ sublattice is naturally different from that of the Mn$^{4+}$ sublattice, 
such difference causes nonlinearity of the $1/\chi(T)$ curve. 
This could be a reason why the deviation appears at temperatures below 150~K. 
Above 200~K, the difference of $C$ and $\Theta$ between two sublattices becomes relatively small compared with temperature, 
resulting in the linear relationship between $1/\chi$ and temperature. 

In order to understand the complex and unique magnetic ground state of \nmo, 
we compare the series of $A$\ce{Mn2O4} ($A$ = Li, Na and Ca). 
Note that the crystal structure of JT-ordered \ce{CaMn2O4} consists of each double-rutile chain, 
which are connected through edge-sharing oxygen atoms. 
This structure is slightly different from the CFO-type structure, in which the chains are interconnected through corner-sharing oxygen atoms instead \cite{Ling2001}. 
Furthermore, the \ce{CaMn2O4} structure has a 180\dd~superexchange path so that the Heisenberg exchange interactions have a priority over the easy axis anisotropy. 
As a result, the spin direction is independent of the orbital ordering pattern (the spins are parallel to the ladders). 
The magnetic ground state of \nmo is rather unique, 
because each of the two different Mn sites has its own ordering temperature and modulation. 
Although no detailed calculation has been reported so far, 
the interaction between Mn$^{3+}$ and Mn$^{4+}$ ladders (in $ab$-plane; $J_4$ and $J_5$) are most likely rather weak compared with the interaction within the each ladders ($J_1$, $J_2$ and $J_3$). 
This is evidenced from the fact that the spin configuration of Mn$^{3+}$ is not affected by the onset of Mn$^{4+}$ spin order. 
Within the Mn$^{3+}$ ladder, 
the single ion anisotropy dominates owing to superimposition of the $3d$ electrons in the $d_{\rm z^2}$ orbitals of Mn$^{3+}$. 
On the other hand, in the Mn$^{4+}$ ladders, 
there is no contribution from anisotropy but the competing exchange interaction is favoured instead, 
because of the absence of the  $d_{\rm z^2}$ orbital order. 
The combination of the complex exchange pathways, i.e. edge and corner-sharing octahedra through the oxygen, 
and geometrical frustration on the triangular lattice often result in a complex incommensurate spin structure, 
such as CFO-\ce{NaV2O4} and \ce{CaCr2O4} \cite{Nozaki2010, Damay2010}.      

As for \lmo, no long-range magnetic order was observed down to 1.5~K by the present neutron powder diffraction experiment. 
This is most likely due to the nonstoichiometry of Li, 
which naturally increases the amount of Mn$^{4+}$ by about 8$\%$ from that (50$\%$) in NaMn$_2$O$_4$. 
As a result, it is impossible to form the ladder occupied only by Mn$^{3+}$ but likely to form the two ladders with random distribution of Mn$^{3+}$ and Mn$^{4+}$, 
which hinders the formation of long-range magnetic order down to 1.5~K.
Therefore, it is concluded that charge and orbital order play a significant role to align the Mn moments in the triangular spin ladders.

This also suggests the possibility that the magnetic ground state of NaMn$_2$O$_4$ is sensitive to the Na content. 
Therefore, it is very interesting to study Na$_{1-\delta}$Mn$_2$O$_4$ prepared by an electrochemical reaction with neutron scattering and $\mu^+$SR.

The spin-orbital coupling in JT active manganites has been reported for various compounds. 
Typical examples are the manganese perovskites, $Ln$\ce{MnO3}, where $Ln$=~La to Gd and Tl \cite{Tokura2000, Khalyavin2016}. 
Such compounds show non-collinear orbital order with collinear spin order, 
owning to the competition between single anisotropy and exchange interactions. 
The compound series of $A_{1-x}A^{'}_{x}$\ce{MnO3} ($A$ = lanthanum, $A'$ = Ca, Sr and Ba) is known to have charge, orbital and magnetic order. 
Similar to \nmo, 
the spin structure of \ce{La_{0.5}Ca_{0.5}MnO3} consists of two different magnetic sublattices, 
which is also based on Mn$^{3+}$ and Mn$^{4+}$ sites \cite{Radaelli1997}. 
Besides the perovskites, very recently, the layered honeycomb \ce{BiMnTeO6} was reported to show the non-collinear spin-orbital coupling \cite{Matsubara2019c}. 

In contrast, layered triangular lattice compounds, like \ce{CuMnO2} and \ce{NaMnO2}, pose a frustrated triangular spin lattice with strong anisotropic interaction due to the ferro-orbital ordering. 
Such magnetic frustration leads to a magnetoelastic coupling, 
evidenced by the structural transition \cite{Damay2009, Giot2007}. 
Moreover, several studies about the correlation between spin and orbital order were reported on \ce{CaMn7O12}, 
which reveal that both crystal (orbital) and magnetic (spin) structures are incommensurate. 
Such unique modulation plays a role in breaking the inversion symmetry, 
resulting in one of the best type-II multiferroic material \cite{Johnson2016, Perks2012}. 
The total understanding based on spin and orbital order would be useful for various technical applications \cite{Vopson2015}.

\nmo~is therefore a novel and intriguing example of a charge-orbital-spin correlated manganites. 
Particularly, inelastic neutron scattering measurements using a single crystal sample together with theoretical work will be extremely valuable to further understand the detailed relationship between orbital physics and magnetism 
not only in manganites but also in general transition metal oxides.

\section{\label{sec:discussion}Conclusions}
In conclusion, this study of CFO-\nmo~has revealed a complex magnetic ground state, based on successive charge, orbital and spin order. 
Based on high-resolution neutron powder diffraction, the Mn$^{3+}$ and Mn$^{4+}$ were fully ordered down to 1.5~K. 
Although 2D short-range spin order presents even at 75~K,   
a commensurate AFM ordered phase appears below 35~K and an incommensurate AFM ordered phase follows below 11~K. 
Surprisingly, the two AFM orders are independent each other. 
The commensurate AFM ordering is based on Mn$^{3+}$ AFM coupling along the chain axis, while the incommensurate AFM ordering is most likely spin-density-wave order of Mn$^{4+}$.
Finally, neutron diffraction studies of the closely related \lmo~compound clarify the absence of long-range order down to 1.5~K. 
The existence of a diffuse scattering below 50~K indicates the presence of short-range magnetic order in \lmo.

\section{\label{sec:discussion}Acknowledgement}
The authors thank H. Nozaki (Toyota Central Research and Development Labs. Inc.) for supporting during the neutron diffraction experiment. The neutron scattering measurements were performed using the DMC and HRPT instruments at the Swiss Spallation Neutron Source (SINQ) of the Paul Scherrer Institute (PSI) in Villigen, Switzerland. The authors wish to thank the staff of PSI for the very valuable help with these experiments. This research is funded by the Swedish Foundation for Strategic Research (SSF) within the Swedish national graduate school in neutron scattering (SwedNess) as well as the Swedish Research Council (VR) through a neutron project grant (BIFROST, Dnr. 2016-06955). Further, Y.S. acknowledge funding from the Swedish Research Council (VR) through a Starting Grant (Dnr. 201705078) as well as Chalmers Area of Advance-Materials Science, 
J.S. was supported by the Ministry of Education, Culture, Sports, Science and Technology (MEXT) of Japan, KAKENHI grant No. 23108003 and Japan Society for the Promotion Science (JSPS) KAKENHI Grant No. JP18H01863, and H.S. acknowledges funding from JSPS KAKENHI Grant No. JP17K05521.


\bibliography{library} 

\end{document}